# Electrostatic control of $Li^+$ density and transport rate in double-gated van der Waals devices

E. Hoenig[1,2], X. Zhang[1,2], C. Li[1,2], J. Tong[1,2], G. Chen[1,2], L. Chen[1,2], D. Domaretskiy[1], D. R. da Costa[3], F. M. Peeters[3,4,5], M. Lozada-Hidalgo[1,2]

[1]Department of Physics and Astronomy, University of Manchester, Manchester, UK
[2]National Graphene Institute, University of Manchester, Manchester, UK
[3]Departamento de Física, Universidade Federal do Ceará, Fortaleza, Brazil
[4]School of Physics and Optoelectronic Engineering, Nanjing University of Information Science and Technology, Nanjing 210044, China.
[5]Departement Fysica, Universiteit Antwerpen, Antwerp, Belgium

**Abstract**

Ion transport in crystalline hosts is controlled by an applied potential that simultaneously sets ionic distribution and transport rate, restricting operation to a one-dimensional control space. Here we show that the transport rate of $Li^+$ ions in double-gated van der Waals devices can be modulated while the system occupies fixed ionic-density states. We measure the ionic current along the van der Waals interfaces between hexagonal boron nitride and graphene or $MoS_2$ while simultaneously monitoring the in-plane electronic response. The ionic current exhibits pronounced hysteresis, with plateaus marking discrete ionic-density states balanced by electronic charge, while an independently tuneable electrochemical-potential drop controls the ionic transport rate. The devices sustain over 1,000 switching cycles and function as hybrid ionic–electronic transistors capable of logic operations and memory retention, with ON/OFF ratios exceeding two orders of magnitude. This work demonstrates a two-dimensional control space for ions intercalated in layered materials.

**Main text**

Electronic transistors enable logic, memory, and signal amplification by independently controlling carrier density and transport. Achieving analogous control with ions remains challenging but is motivated by energy storage[1], biosensing[2] and applications in nanofluidics[3-8] and ionic neuromorphic computing[9-12]. Nanofluidic devices have made important progress toward this goal. When channel widths approach the Debye length of the electrolyte, ionic density becomes dominated by surface charge[13], enabling nanofluidic transistors and memristors based on nanochannels[6], carbon nanotubes[10], and two-dimensional materials[7,8,11,12]. A different approach is to control ions within a solid host rather than in a confined electrolyte. In layered materials, ions (e.g. protons[14-20] or lithium ions[21-27]) can intercalate in the crystal lattice at characteristic electrochemical potentials[21,28], enabling energy storage in batteries or modulation of the magnetic or electronic properties in the host. However, this potential simultaneously controls the transport kinetics[1,29-31], restricting ion transport to a one-dimensional operating space. Here we expand the control space for $Li^+$ transport in van der Waals devices using a technique known as double gating. Originally developed to control electronic phases in 2D devices (e.g., Mott insulating, superconducting), this technique decouples the charge density and electric field in the devices through two independent gate voltages[32-39]. Applied to Li transport in van der Waals devices, this geometry introduces two independent controls at the van der Waals interface, allowing ionic occupancy and transport rate to be varied separately. This enables direct mapping of $Li^+$ transport as a function of ionic density and electrochemical potential drop along the channel and allows transport to be accelerated without necessarily increasing the potential of the ion-supplying reservoir.

**Results**

Our devices consist of mechanically exfoliated few-layer graphene (FLG) or $MoS_2$, stacked on a <50 nm thick hexagonal boron nitride (hBN) flake with a circular hole 3–5 μm in diameter (Fig. 1a, Fig. S1). The 2D stack was supported on a silicon nitride substrate with an aperture aligned with the hole in the hBN crystal (Fig. S1). Both sides of the devices were coated with a nonaqueous lithium-conducting electrolyte, lithium bis(trifluoromethanesulfonyl)imide (LiTFSI) dissolved in polyethylene glycol (PEG)[39] and contacted with two centimetre-sized carbon cloth gate electrodes. The FLG or $MoS_2$ flakes were patterned such that the van der Waals interface between hBN and the 2D crystals effectively formed a radially symmetric channel (typical length $L \approx 5$ μm) connecting both electrolyte reservoirs (Fig. 1a, Fig. S1). This channel is the only pathway for ionic transport between reservoirs, since pristine graphene and $MoS_2$ are impermeable to $Li^+$ ions through their basal planes[39,41,42] (Fig. 1b). For measurements, devices were placed in an inert argon atmosphere and heated to 50 °C to facilitate ion intercalation. The FLG or $MoS_2$ flakes were connected to the electrical circuit shown schematically in Fig. 1a. Two gate voltages, $V_t$ and $V_b$ (top and bottom, respectively), were applied to the devices. This enabled independent control of the potential on each device-electrolyte interface, which was monitored with a Pt wire pseudo-reference electrode (Fig. S2). An additional drain-source bias, $V_{ds}$, was used to measure the in-plane electronic conductivity. The devices thus enable measurement of both $Li^+$ transport through the van der Waals interface and the in-plane electronic current in the crystal. We set the control variables as the sum and difference of the gate voltages, $\Sigma V = V_b + V_t$ and $\Delta V = V_b - V_t$, following the standard protocol for double-gated devices[33]. We adopt this convention and later show that these two variables control the charge density in the device and the electrochemical potential drop of $Li^+$ ions along the channel, respectively.

Fig. 1b shows the typical ionic current response, $i$, of FLG and $MoS_2$ devices at fixed $\Delta V$ for $\Sigma V > 0$, such that the devices were electron doped. At low $\Sigma V$, the current remained within the background level (<10 pA), but beyond a threshold, which depended on the crystal (FLG or $MoS_2$), it rose sharply, sometimes displaying closely spaced jumps, before reaching a plateau. On the reverse sweep, the current remained high but decreased in broad and well-defined steps, until a second threshold was reached and abruptly returned to the baseline. These features were reproducible across multiple cycles of a device and across 14 different devices (Fig. S3, Fig. S4). The magnitude and qualitative form of the ionic-current response were also preserved when scanning rate was varied by more than an order of magnitude (Fig. S11). Overall, the ionic response displayed a hysteresis loop with an ON-OFF ratio of $\approx 10^2$, and devices could sustain >1,000 of such ON-OFF cycles without degradation (Fig. 1e). To confirm that the transport occurs along the 2D interface, we measured reference devices with channel edges blocked by the SU-8 seal. These showed no measurable current within our experimental background (grey line, Fig. 1b), as expected from these crystals' impermeability to $Li^+$[39,42]. Additional control experiments using either $EMIM^+$ ions instead of $Li^+$ or sweeping $\Sigma V$ into the hole-doping regime (negative voltages) to intercalate $TFSI^-$ ions yielded irreversible leakage and device failure, confirming that the phenomena observed here originate from $Li^+$ transport along the 2D interface.

Fig. 1c shows how the ionic response depends on $\Delta V$. While the $i$-$\Sigma V$ traces remained qualitatively similar, the current increased notably with $\Delta V$. To quantify this dependence, we recorded $i$ at a fixed value of $\Sigma V$ at mid-plateau (red circles, inset Fig. 1c), for different values of $\Delta V$. This revealed a linear $i$–$\Delta V$ relation for the plateau current. However, this was only observed when the system was ON. Stepping $\Delta V$ at the same $\Sigma V$ without first crossing the ON threshold, such that the system was OFF, kept $i$ at the baseline. This demonstrates that $\Delta V$ linearly increases the ionic current through the device, but only if $\Sigma V$ sets the channel to its open state. For reference, we measured our devices in a single-gated configuration. In this case, the large hysteresis disappeared, and the devices exhibited irreversible leakage after ~10 ON–OFF cycles (Fig. S5), demonstrating that double gating is essential for our observations.

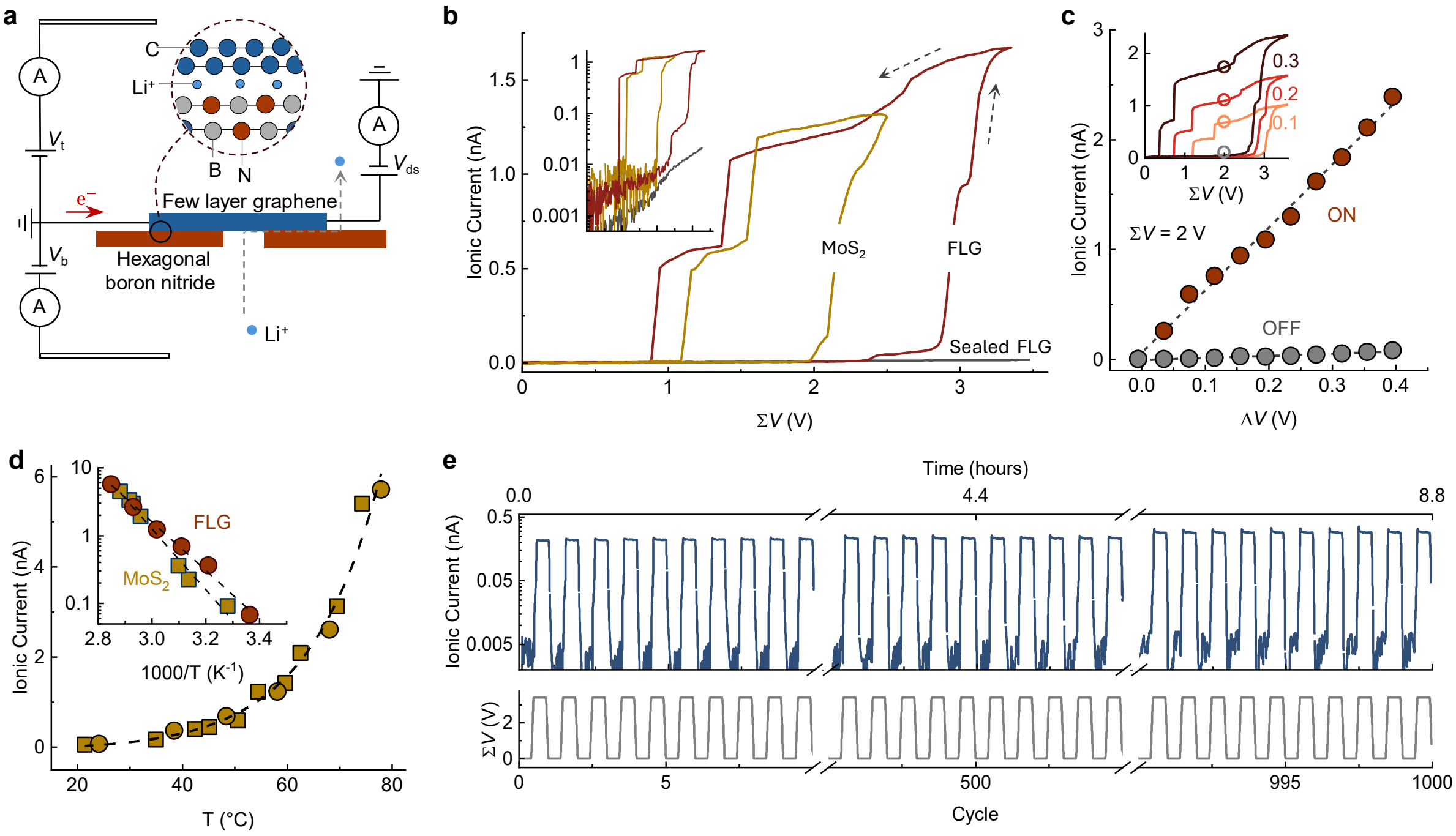

**Fig. 1. Control of $Li^+$ transport in double-gated FLG/$MoS_2$ devices. a**, Schematic of few-layer graphene (FLG) devices. **b**, Charge-discharge curves as a function of $\Sigma V = V_b + V_t$ for FLG (red), $MoS_2$ (yellow) and sealed FLG (grey) devices under constant $\Delta V = V_b - V_t = 0.2$ V for FLG and sealed FLG and $\Delta V = 0.4$ V for $MoS_2$. Arrows indicate the scanning direction. Inset, data from the main panel in log scale. **c**, $\Delta V$ dependence of ionic current for a typical FLG device under constant $\Sigma V = 2$ V, with the different data sets corresponding to ON (red) and OFF (grey). Grey dashed lines, linear fits to data. Inset, charge-discharge curves for $\Delta V = 0.1$-$0.3$ V. Red and grey circles mark current at $\Sigma V = 2$ V for the different traces plotted in the main panel. **d**, Temperature dependence of ion transport current for two $MoS_2$ devices, represented by gold circle and square symbols, respectively. Each data point is the average current at the highest current plateau. Dashed black line, exponential fit to the data. Inset, Arrhenius plots of FLG (red) and $MoS_2$ (gold) devices. Dashed lines, best fits to data. **e**, Long-term cycle stability test. The device is switched between ON ($\Sigma V = 3.2$ V) and OFF ($\Sigma V = 0$ V) states at constant $\Delta V = 0.2$ V at a frequency of 30 mHz for 1,000 cycles, corresponding to nearly 9 hours of continuous operation.

To gain further insights into the ionic transport, we measured the temperature dependence of *i* at the highest plateau. In the OFF state, only background levels of current were detected within the temperature range tested (20-80 °C). In contrast, in the ON state, the plateau current increased exponentially with temperature, rising by approximately two orders of magnitude across this interval, indicating activated transport (Fig. 1d). Arrhenius plots from multiple devices revealed that the activation energy of the process was $E_a$ = 0.6 ± 0.2 eV for FLG and 0.7 ± 0.1 eV for $MoS_2$ devices (Fig. 1d inset, Fig. S6). Reference devices without 2D crystals (open-hole devices) displayed current with linear temperature dependence that increased only ~5-fold over the same temperature range (Fig. S7), confirming that the $E_a$ measured in our devices is intrinsic to $Li^+$ entry into the 2D crystal/hBN interface. Together, these results show that $\Sigma V$ switches $Li^+$ transport between OFF and a thermally activated ON state, while $\Delta V$ independently modulates the current.

We then analysed the electronic and ionic responses together, focusing on FLG. Starting with the electronic response, Fig. 2a shows that during the intercalation sweep (orange dashed curve), the electronic current, *I*, displayed a minimum at $\Sigma V \approx 0.2$ V, corresponding to the charge neutrality point of FLG. The current then increased smoothly with $\Sigma V$ until it displayed a sharp step increase marked as '0' in Fig. 2a. This event could be clearly isolated in the differential conductance trace, $dI/d(\Sigma V)$, as shown in Fig. 2b, where it appears as a sharp peak in the intercalation sweep (orange dashed curve). This trace also revealed three additional peaks (labelled 1–3) in the reverse, deintercalation, sweep (blue curve). Maps of the differential conductance in the full $\Sigma V$–$\Delta V$ space (Fig. 2c) revealed that all these peaks persisted across the $\Delta V$ range tested, forming continuous features in the maps. Notably, numerical analysis (Fig. 2d) revealed that the charge associated with the intercalation peak, $\Delta n_0$, is approximately equal to the total charge in the deintercalation peaks, $\Delta n_0 \approx \Delta n_1 + \Delta n_2 + \Delta n_3$, and that the charge in all these peaks is independent of $\Delta V$ (Fig. S8 and Section S4). These findings demonstrate that the electronic signal exhibits balanced charge–discharge events during the intercalation–deintercalation cycle controlled by ΣV that are independent of $\Delta V$.

Further insight comes from the ionic current maps. The intercalation map (Fig. 2f, left) shows a single linear feature marking the onset of intercalation. This feature coincides with the main charging peak ($\Delta n_0$) in the corresponding electronic transport map and occurs at a constant back-gate voltage of $V_b$ = 1.5 ± 0.2 V ($V_b$ = 1.2± 0.2 V for $MoS_2$), consistent with the intercalation potential reported for comparable single-gated devices (Fig. S3) [15,43]. By contrast, the deintercalation map (Fig. 2f, right) displays three distinct nonlinear features. These features are quantitatively reproducible across cycles, qualitatively consistent across devices (Fig. S4), and coincide precisely with the nonlinear features in the deintercalation electronic transport map. The corresponding ionic current traces (Fig. 2e) show that these features mark the onset of the broad ionic current plateaus discussed in Fig. 1 (labels 0–3). Hence, within each plateau, the electronic charge density and ionic current remain constant, reflecting ionic-electronic charge balance during $Li^+$ intercalation.

This balance allows for an estimate of the ionic density from the electronic charge-discharge analysis in Fig. 2d. This reveals $n_{Li} \approx 6\text{x}10^{13}$ cm$^{-2}$ at the highest plateau, decreasing in steps of $\Delta n_{Li} \approx 2\text{x}10^{13}$ cm$^{-2}$ until the interface is discharged (Section S4). This, in turn, reveals how $\Delta V$ modulates the ionic current. Since the ionic density is fixed at each plateau, we conclude that $\Delta V$ controls the driving force for $Li^+$ ions; namely, the electrochemical potential drop along the channel, $\Delta\tilde{\mu} = e\,\Delta V$. The acceleration of the transport is then quantified by analysing the average current in each of the three plateaus as a function of $\Delta V$ (Fig. 2g). This shows that the ionic current increases linearly with $\Delta V$, with different slopes for each plateau, reflecting their different $n_{Li}$. These observations are consistent with the well-established principle that ionic transport is driven by the gradient of the electrochemical potential. In our system, this gradient is directly controlled by the potential difference between the channel ends, which can be set independently of ionic density.

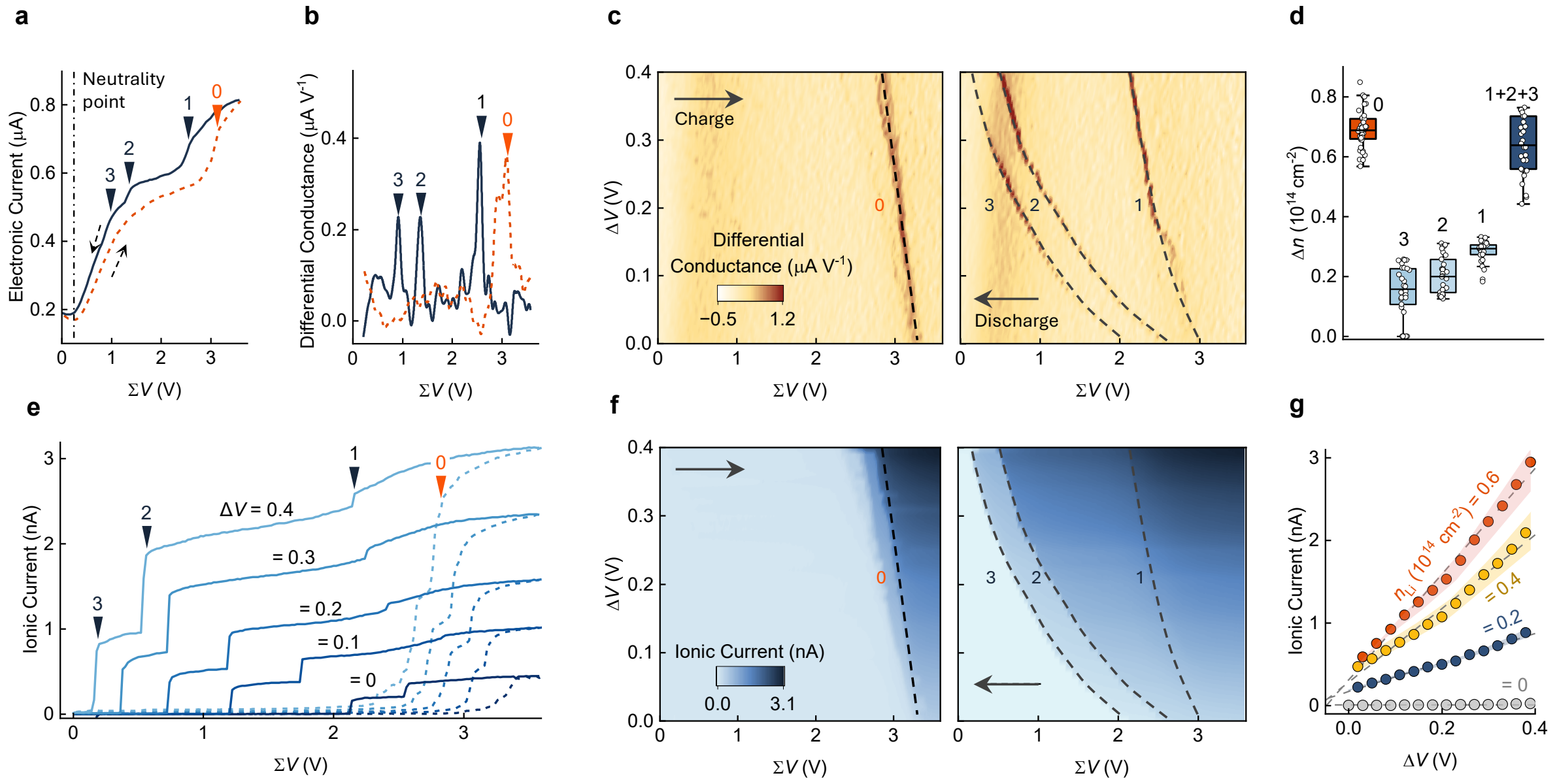

**Fig. 2. Maps of $Li^+$ and electronic transport for FLG devices.** **a**, Electronic current, $I$, versus $\Sigma V = V_b + V_t$ at constant $\Delta V = V_b - V_t = 0.2$ V. Orange and blue triangles, charge and discharge events, respectively; shown as peaks in panels **b** and **c**. Vertical line, position of the neutrality point (NP). Dashed orange and solid blue curves, intercalation (forward) and deintercalation (backward) sweeps. **b**, Background-subtracted electronic differential conductance, $dI/d(\Sigma V)$, obtained from panel **a**. Each peak corresponds to an electronic charge–discharge event, and the area under the peak quantifies the event's charge, $\Delta n$. **c**, Maps of electronic differential conductance vs. $\Sigma V$ and $\Delta V$. Left, intercalation ($\Sigma V$ swept from low to high). Right, deintercalation ($\Sigma V$ swept from high to low). Arrows indicate sweep direction. Dashed lines, guides to the eye. **d**, Box plot of $\Delta n$ for charge-discharge events, labelled 0-3. Individual data points show $\Delta n$ extracted for different $\Delta V$. Data set labelled 1+2+3 corresponds to the sum of discharge peaks 1-3 (Section S4). **e**, Ionic transport traces for various constant $\Delta V$ obtained from panel **f**. The traces show plateaus, at which the ionic density, $n_{Li}$, is constant. Dashed and solid curves correspond to forward and backward sweeps, respectively. **f**, Map of $Li^+$ transport current as a function of $\Sigma V$ and $\Delta V$, measured simultaneously with in-plane electronic current. Dashed lines, guides to the eye. **g**, Ionic current at plateaus as a function of $\Delta V$. The $Li^+$ density ($n_{Li}$) for each plateau is indicated. Points show the average plateau current, and the shaded regions denote the min–max range for each plateau. Grey dashed lines are linear fits, demonstrating that $i \propto n_{Li}\,\Delta V$ for fixed plateau (constant $n_{Li}$).

These results provide a complete picture of $Li^+$ transport in double-gated devices. The $Li^+$ density in each plateau, $n_{Li}$, is fixed and balanced by the electronic charge and their dependence in the applied gate voltages can be explained analytically (Fig. S10, Section S7). The resulting ionic current, $i$, on the other hand, is proportional to the difference in the electrochemical potential defined by the two gates, $\Delta\tilde{\mu}$. This allows describing the transport phenomenologically using a drift–diffusion model along a 2D channel: $i = A\, n_{Li}\, \mu\, \Delta\tilde{\mu}$, where $A$ is a unitless device geometry constant (Section S5) and $\mu$ the effective mobility of $Li^+$. The latter term incorporates a thermally activated factor, $\exp(-E_a/kT)$, where $E_a = 0.6 \pm 0.2$ eV, reflects the barrier for $Li^+$ entry into the channel. An order-of-magnitude estimate of the effective mobility can thus be derived from the transport data as: $\mu = (A\ e\ n_{Li})^{-1}\ \partial i/\partial \Delta V \approx 1\times10^{-4}\ cm^2\ V^{-1}\ s^{-1}$ for all plateaus. We find that the estimated $Li^+$ density and mobility are in reasonable agreement with previous reports for $Li^+$ transport in 2D devices[15,28,32] (Section S5). Taken together, these observations demonstrate that double-gated devices enable transistor-like control of $Li^+$ transport in van der Waals devices.

Double gating offers two key performance advantages over single-voltage controlled architectures. First, ON-OFF cyclability is greatly enhanced. Whereas single-voltage controlled 2D devices can typically be measured for up to ~10 cycles[15,28,29,32,43] (including our own, Fig. S5), double-gated devices can sustain >1,000 cycles without degradation. Second, double gating enables control of transport by tuning the electrochemical potential drop along the channel, rather than relying solely on a single applied potential. This allows acceleration of the transport without necessarily increasing the bias at the electrolyte reservoir ($V_b$) that supplies the ions. In single-gate controlled devices, faster ion transport often requires higher operating potentials, which can induce parasitic side reactions or even device failure[23]. Our findings suggest that double gating could enable future charge storage architectures that exploit this new way of accelerating ionic transport.

To illustrate the robust control over intercalation enabled by our devices, we operate them as information-processing elements (Fig. 3a). In the first mode, they function as ionic transistors. The ionic current is switched ON or OFF via $\Sigma V$, which functions as a digital WRITE signal, while $\Delta V$ continuously tunes the current magnitude, functioning as an analogue READ signal[44]. Fig. 3b shows the typical current response when a 50 mV sine wave READ is overlaid with a square-wave WRITE signal. When ON (high $\Sigma V$), the sine wave yields an AC transport signal and when OFF, no current is detected within the experimental background. We found that this mode can also be employed to isolate small ionic signals. The AC modulation described above drives a $Li^+$ transport current, which enables selecting the response only at the set frequency, thus enabling its selective extraction, similar to the operation of a lock-in amplifier.

In the second operation mode, we exploit the hysteresis introduced during charge–discharge cycles to construct programmable memory devices with FLG and $MoS_2$. To this end, $\Sigma V$ was held between the intercalation and deintercalation thresholds, with $\Delta V$ fixed. A short WRITE pulse with $\Sigma V$ above the intercalation threshold set the device to the ON state (Fig. 3c, d), with current 100 times larger than the background. Because the channel remains charged after the pulse, the system retains this ON state until a second pulse with $\Sigma V$ below the deintercalation threshold switches the device OFF. The ionic current thus exhibits non-volatile memory retention, with FLG and $MoS_2$ exhibiting qualitatively equivalent signals. The associated electronic responses provide a secondary signal that tracks the intercalation–deintercalation events, superimposed with a volatile component that follows the WRITE voltage pulses. In $MoS_2$, this volatile response drives the device fully into an insulating state during the OFF pulse. Our 2D $Li^+$ devices thus integrate non-volatile ionic memory and volatile electronic responses in a new platform for iontronic applications[11,12,24,25,45].

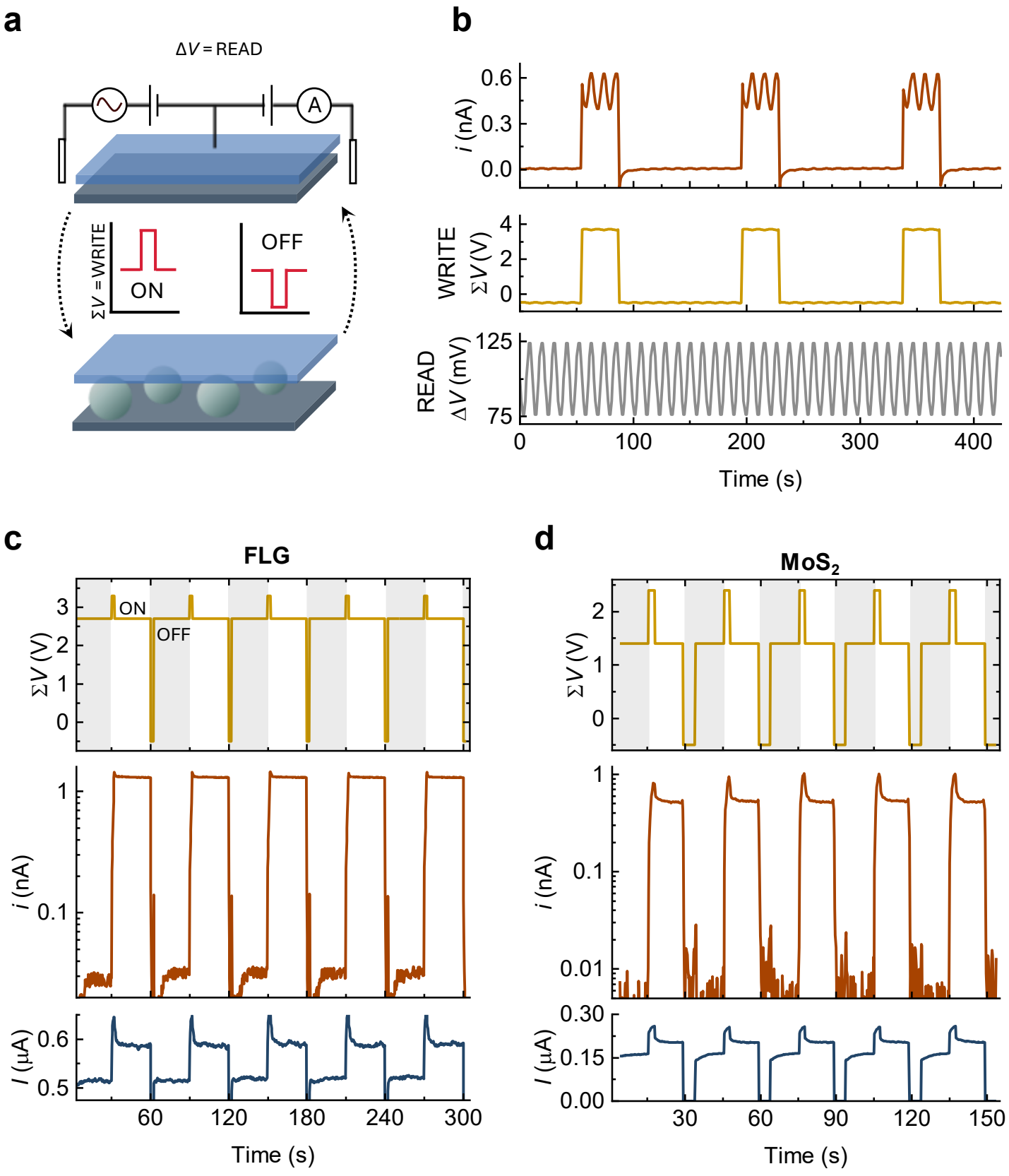

**Fig. 3. Independent control of charge and electric field enables precise logic and memory devices. a**, Schematic of the signal protocol. Σ$V$ and Δ$V$ are set as the digital WRITE and analogue READ inputs, respectively. The schematic shows that the system is ON (OFF) when a high (low) pulse is applied to Σ$V$, charging (discharging) $Li^+$ ions (shown as grey spheres) in the hBN/FLG interface (shown as planes). A small AC bias (50 mV peak-to-peak) is applied to the top gate to continuously tune the READ input. A small in-plane bias, $V_{ds}$, is applied to readout electronic conductance, with the source connected to ground. **b**, Combined WRITE-READ process. A small AC bias is applied to the top gate, varying the READ input, while a square wave is applied to the WRITE input, periodically switching between high and low states. Top panel (red traces), ionic current, $i$, response. Middle (bottom) panel is the voltage waveform applied to Σ$V$ (Δ$V$). **c-d**, Demonstration of logic and memory operations on FLG (**c**) and $MoS_2$ (**d**) devices. We hold READ constant and apply high and low WRITE pulses to switch the state of devices. Top panels (yellow traces) are voltage waveforms applied to Σ$V$ for FLG (**c**) and $MoS_2$ (**d**). Middle panels (red traces) are corresponding ionic current, $i$, responses, and bottom panels (blue traces) show the responses of in-plane electronic current, $I$.

### Discussion

We demonstrated transistor-like control of ionic transport in double-gated van der Waals devices by decoupling $Li^+$ density and transport rate. Compared to similar single-gated architectures using the same materials, this approach significantly improves cycling endurance and enables faster ion transport without increasing the potential of the electrolyte supplying the ions. This finding may inform protocols to accelerate interfacial transport while mitigating high-potential parasitic reactions in battery-relevant interfaces. This approach could also be explored with multivalent charge carriers like $Mg^{2+}$ and $Zn^{2+}$, which are often limited by slow interfacial kinetics[46]. More broadly, this work expands the control parameter space during ion intercalation into host crystals, suggesting new design possibilities for charge storage and programmable ionic memory devices.

**Supporting Information**

Methods (sections S1 to S7) including device fabrication, measurement protocols and theoretical description of $Li^+$ transport. Figures S1 to S12, showing device schematics and images, additional device and electrolyte characterization, reproducibility and theoretical results (PDF).

Supporting Information for

# Electrostatic control of $Li^+$ density and transport rate in double-gated van der Waals devices

E. Hoenig[1,2], X. Zhang[1,2], C. Li[1,2], J. Tong[1,2], G. Chen[1,2], L. Chen[1,2], D. Domaretskiy[1], D. R. da Costa[3], F. M. Peeters[3,4,5], M. Lozada-Hidalgo[1,2]

[1]Department of Physics and Astronomy, University of Manchester, Manchester, UK
[2]National Graphene Institute, University of Manchester, Manchester, UK
[3]Departamento de Física, Universidade Federal do Ceará, Fortaleza, Brazil
[4]School of Physics and Optoelectronic Engineering, Nanjing University of Information Science and Technology, Nanjing 210044, China.
[5]Departement Fysica, Universiteit Antwerpen, Antwerp, Belgium

**Section S1: Device Fabrication**

Devices were fabricated on silicon substrates coated on both sides with a 500 nm SiNx layer. Apertures 3-5 μm in diameter were defined by photolithography, wet etching, and reactive ion etching (RIE) [1]. Source and drain electrodes (Cr/Au) were deposited via photolithography and electron-beam evaporation. Hexagonal boron nitride (hBN, <50 nm thick) was mechanically exfoliated and transferred onto the substrate using polypropylene carbonate dry transfer[2]. Then, apertures in the hBN were patterned by RIE with the SiN substrate acting as a hard mask and then cleaned with mild RIE. We found that this cleaning step is important to ensure device robustness. Few-layer graphene (FLG) or $MoS_2$ layers were exfoliated, patterned by photolithography and RIE, and transferred onto the hBN/SiNx substrate to cover the aperture. For $MoS_2$ devices, a graphite electrode (<10 layers) was patterned and transferred onto the $MoS_2$ flake to electrically connect the $MoS_2$ flake to the Au electrodes. The stacks were sealed with pre-patterned SU-8 clamps[3]. The device was then mounted on PCB boards and transferred to an argon glovebox, where it was coated on both sides with approximately 15 μL of lithium-conducting electrolyte. The devices were electrically contacted with two centimetre-sized carbon cloth sheets as gate electrodes and Pt wire as the pseudo-reference electrode. The electrolyte used was 1:10 w/w LiTFSI in polyethylene glycol 300 (PEG 300) prepared by drying LiTFSI at 70 °C overnight and PEG 300 at 60 °C under vacuum with stirring in an inert atmosphere. All measurements were performed in a temperature-controlled, sealed chamber with inert argon atmosphere. The temperature was kept at 50 °C, except for *T*-dependence measurements.

**Section S2: Transport Measurements**

For electrical measurements, a dual-channel Keithley 2636B source-meter was used to apply top and bottom gate voltages. The two gate potentials result in two $Li^+$ current signals, top ($i_t$) and bottom ($i_b$) gate current, which characterise the two halves of the lithium transport circuit, namely $Li^+$ transport from one gate electrode towards the heterostructure, and then from the heterostructure to the other gate. We find that these two signals are nearly identical in magnitude, differing by less than 5% of the $Li^+$ transport signal. For this reason, it is sufficient to use only one signal to unambiguously characterise steady-state $Li^+$ transport, similar to our previous work on double-gated proton transport through graphene[4]. A second Keithley 2614 source-meter was used to apply a 1 mV and 50 mV drain-source bias ($V_{ds}$) to FLG and $MoS_2$, respectively, and measure the in-plane electronic current, *I*. The stability of the Pt pseudo-reference was confirmed using the graphene Dirac point as an internal electronic reference. The measured Dirac-point voltage varied by only ~100 mV across devices and over the timescale of full gate maps, which we take as the uncertainty in the absolute potential scale. This uncertainty is smaller than the device-to-device scatter in the extracted intercalation thresholds and does not affect the differential bias $\Delta V = V_b - V_t$.

We measured electronic current ($I$) and ionic current ($i$) simultaneously as a function of $\Sigma V = V_b + V_t$ and $\Delta V = V_b - V_t$ using a custom python code that allowed controlling $\Sigma V$ and $\Delta V$ as independent variables. We confirmed that both gate potentials are independent by monitoring the top-gate potential ($V_t$) with a reference electrode while sweeping the bottom gate ($V_b$), and vice versa (Fig. S2). The measured reference voltage ($V_t^{ref}$) remained constant within 10 mV when $V_b$ was varied by > 2 V, confirming independent gate operation.

Before measurements, we perform several intercalation-deintercalation sweeps until the intercalation response stabilises. To obtain maps of $I$ and $i$, we swept $\Sigma V$ for a fixed $\Delta V$ at a rate of 10 mV $s^{-1}$ and stepped $\Delta V$ at intervals of 10 mV. The maximum sweep range for $\Sigma V$ is -0.5 to 4.0V and $\Delta V$ is 0 to 0.5 V. We typically avoided applying biases beyond this range to prevent damage to the devices. For temperature-dependence measurements, the temperature was set to the target temperature and once stabilised, $\Sigma V$ was swept to record a full charge-discharge curves at fixed $\Delta V$ = 0.1 V.

**Section S3: Electrolyte Characterisation**

To confirm that the bulk electrolyte does not limit ion transport in our devices, we measured devices consisting only of a $SiN_x$ substrate with an aperture ('open hole device'). Fig. S7 shows that the $I$–$V$ characteristics of these open-hole devices were linear across the range of temperatures used in our measurements. The current was several orders of magnitude higher than that measured in our system, demonstrating that the effects observed in our devices arise from the FLG/$MoS_2$ channel, rather than from the electrolyte.

The electrolyte capacitance was measured with devices consisting of two gold electrodes patterned on $SiO_2$ masked to expose only a set active area. This area differed by >100×, which ensured the smaller electrode dominated the measured capacitance. Cyclic voltammetry (CV) was performed from −0.1 V to 0.1 V at scan rates of 5–80 mV $s^{-1}$. Fig. S9 shows that the CV curves displayed no redox peaks or asymmetry between the positive and negative voltage branches. The area-normalised capacitance of the electrolyte, $C$, could then be obtained from the CV curves from the expression $C = (A \times \Delta V \times v)^{-1} \int I \, dV$, where $A$ is the active area of the electrode, $\Delta V$ is the voltage range in the CV, $I$ is the current and $v$ is the scan speed. For the scan rates used in this work (10 mV $s^{-1}$), $C \approx 15$ μF $cm^{-2}$, which we use in our estimates involving $C$.

**Section S4: Decoupled charge density and ionic transport**

The electronic charge density at the FLG/hBN interface has two contributions: (i) smooth electrostatic doping and (ii) charge introduced in stepwise intercalation events. In our devices, only the top gate capacitively couples to the majority of the FLG/hBN interface, so the electrostatically induced charge is determined by $V_t$ as: $n = C e^{-1} (V_t - \Delta_{NP})$, where $C = 15$ μF $cm^{-2}$ is the electrolyte capacitance (Fig. S9), $e$ the elementary charge and $\Delta_{NP}$ is the charge neutrality point of FLG. $C$ is dominated by the double layer capacitance, due to the relatively large quantum capacitance of few layer graphene (>30 μF $cm^{-2}$ for 6 layers)[5].

To quantify the electronic density associated with these features, we recast the measured dI/d(ΣV) signal in terms of electrostatically induced charge by converting it to $dI/dn$ using the gate-induced carrier density relation given above. In this form, smooth variations arise from capacitive gating, whereas the observed sharp features directly reflect charge transferred during discrete intercalation events, enabling their quantification. We then integrate the background-subtracted $dI/dn$ signal over a given peak: $\Delta I = \int (dI/dn) \, dn$, which gives the total current change associated with that event. Because in-plane electronic current scales linearly with carrier density ($I \propto n$), $\Delta I$ is proportional to the injected carrier density: $\Delta I = \beta \, \Delta n$, with $\beta$ estimated from the slope of the $I(n)$ relation near the neutrality point: $\beta = (dI/dn)|_{NP}$. The electronic density added in each charging event is then estimated as: $\Delta n \approx \beta$-1 $\int (dI/dn) \, dn$. The $Li^+$ density is estimated from these discrete charging events. Prior theoretical calculations and experimental studies show that each intercalated $Li^+$ is approximately charge-compensated by one electron ($\Delta n_{\mathrm{Li}} = \alpha\, \Delta n$, with $\alpha$~0.8-1)5-8. In this work, we take $\alpha = 1$, giving $\Delta n_{\mathrm{Li}} \approx \Delta n$.

We verify that $\Delta V$ controls transport, and $\Sigma V$ controls charge density by quantitatively deriving effect-sizes of both parameters on ionic and electronic signals. For each ionic density plateau, we evaluated how the average ionic current depends on $\Delta V$. In all cases the current varies linearly with $\Delta V$, with slopes significantly different from zero (Fig. 2g): Stage 1, m = 1.75 ± 0.05 nA/V (p = $3\times10^{-12}$, $R^2$ = 0.99); Stage 2, m = 4.4 ± 0.2 nA/V (p = $3.4\times10^{-11}$, $R^2$ = 0.98); Stage 3, m = 6.4 ± 0.2 nA/V (p = $1.9\times10^{-12}$, $R^2$ = 0.99). This confirms that $\Delta V$ linearly sets the transport rate at each fixed density stage. We get complementary evidence from the charge density calculated using the electronic response. The deintercalation sum is accounted for by the intercalation charge to within 10%, $\Delta n_1+\Delta n_2+\Delta n_3$ = 0.63±0.15 vs $\Delta n_0$ = 0.69±0.06, and a linear fit to the total charge (Fig. S8) produces a slope not significantly different from zero ($0.06\times10^{14}$ $\mathrm{cm^{-2}\,V^{-1}}$, $p$=0.39, $R^2$=0.019). The consistent charge balance ($\Delta n_0 \approx \Delta n_1+\Delta n_2+\Delta n_3$), and the independence of the total charge across the full $\Delta V$ range tested indicates that, within our resolution, the total $Li^+$ occupancy set at each $\Sigma V$-defined stage is not measurably altered by $\Delta V$.

Our assignment of the plateaus to discrete $Li^+$-occupied states is based on two key observations: first, the plateau transitions are accompanied by sharp changes in the electronic conductance of the host crystal, and that the plateau current within a given state scales approximately linearly with $\Delta V$, consistent with field-driven transport through a channel whose $Li^+$ occupancy is fixed. However, the present measurements do not uniquely identify the microscopic mechanism. One possibility is the existence of spatial domains. Another possibility is ordered lithium phases or in-plane staging, analogous to those reported in few-layer graphene systems[6]. We rule out interfacial redox states as the mechanism; redox reactions produce currents of the same sign at both gates corresponding to charge injection/removal symmetrically at both gates. In contrast, we observe equal-magnitude opposite-sign signals, consistent with net charge transfer between the two reservoirs. Moreover, the plateau currents scale with $\Delta V$, as expected for field-driven transport, rather than appearing only at a fixed redox potential. We also rule out trap filling, which is a quasi-continuous process and would produce a gradual accumulation of charge, rather than discrete plateaus[7,8].

**Section S5: Estimation of effective $Li^+$ mobility**

Ion transport through the interfacial channel is modelled as two-dimensional radial drift between the inner SiN aperture and the outer FLG edge. The channel therefore has a Corbino-like geometry, with inner radius $r_1$ = 1.5 μm and outer radius $r_2$ = 7.5 μm. For a $Li^+$ density $n_{\mathrm{Li}}$ and mobility $\mu$, the two-dimensional ionic current density is

$$j_r = e n_{\mathrm{Li}} \mu E_r,$$

where $j_r$ has units of current per length, $E_R$ is the radial electric field, and $e$ is the elementary charge. In radial geometry, $E_r = -\frac{dV}{dr}$. The total current crossing a circle of radius (r) is therefore

$$i = 2\pi r j_r = -2\pi r e n_{\mathrm{Li}}\mu \frac{dV}{dr}.$$

At steady state, $i$ is independent of $r$, so we can rearrange and integrate from $r_1$ to $r_2$: $dV = -\frac{i}{2\pi e n_{\mathrm{Li}}\mu}\frac{dr}{r}$, $\Delta V = \frac{i}{2\pi e n_{\mathrm{Li}}\mu}\ln\left(\frac{r_2}{r_1}\right)$. Solving for the current gives

$$i = \frac{2\pi}{\ln(r_2/r_1)} e n_{\mathrm{Li}}\mu\Delta V.$$

Thus, the geometric proportionality factor is $A = \frac{2\pi}{\ln(r_2/r_1)}$, and the ionic current can therefore be written compactly as $i = Aen_{\mathrm{Li}}\mu\Delta V$.

**Section S6: Logic and cyclability tests**

For logic and memory measurements, we define $\Sigma V = V_b + V_t$ and $\Delta V = V_b - V_t$ as the WRITE and READ signals, respectively. The Li transport current (gate current), $i$, and electronic current, $I$, are the two signal readouts. $\Sigma V$ controls whether the ionic current is ON ($\Sigma V$ is HIGH) or OFF ($\Sigma V$ is LOW). Because of the hysteretic response of the device, $\Sigma V$ can also set the device into an intermediate, history dependent state ($\Sigma V$ is set at an intermediate value between HIGH and LOW).

To test the endurance of the devices, we switch $\Sigma V$ between 0 V and 3.2 V to erase or write the signal in our logic device and keep $\Delta V$ = 0.2 V to monitor the device state. We find that the ON/OFF traces were qualitatively independent of frequency when switched between 30 and 300 mHz. For higher frequencies, the response becomes limited by the kinetics of intercalation processes and capacitive transients.

For analogue modulation of the ionic response, the bottom gate, $V_b$ was fixed and a small-amplitude, low-frequency, AC bias modulation voltage, $V_m$ = 50 mV, was superimposed to the top gate voltage, $V_t$. This modulation voltage is about two orders of magnitude less than that needed to switch between ON/OFF states. This mode can also be used to measure the devices with even greater precision, by using the modulation voltage similar to the AC voltage in a lock-in amplifier. This is achieved by sweeping $\Sigma V$ and $\Delta V$ as before for a fixed modulation voltage, $V_m$. The resulting sinusoidal response provides the ionic conductance. This signal is independent of any background current and allows reconstructing the $i$-$\Sigma V$ curve, similar to a lock-in amplifier. Although it is not necessary to apply this technique for the devices in this work, it could be useful for other systems with lower signal-to-noise ratios.

To investigate memory retention in the devices, the WRITE signal was used to set the devices into ON (high pulsing state), OFF (low pulsing state) and HOLD (retention state) by applying $\Sigma V$ = 4.0 V, -0.5 V and 2.0 V, respectively. We keep READ to 0.2 V to record the state of the device. During the measurement protocol, the device is set to the HOLD state, and then we applied short pulses (<1 s) to set it to ON or OFF periodically.

**Section S7: Theory**

The 2D channel is modelled as a negatively charged (electron density $n$) graphene layer separated by a distance $d$ from bulk hBN (dielectric constant $\varepsilon_{hBN} \cong 3.4\, \varepsilon_0$ and $\varepsilon_0$ the vacuum permittivity constant). $Li^+$ ions form a layer with density $n_{\mathrm{Li}}$ in the channel. The Gibbs free energy per unit area of the intercalated system is:

$$G = E_{\mathrm{Li-Li}} + E_{\mathrm{Li-host}} + n_{\mathrm{Li}}(a_{\mathrm{Born}} - eV_{\mathrm{b}}). \qquad (1)$$

Here, $E_{\mathrm{Li-Li}}$ is the interaction energy between $Li^+$ ions, $E_{\mathrm{Li-host}}$ is the interaction energy between $Li^+$ ions and the host and $n_{\mathrm{Li}}(a_{\mathrm{Born}} - eV_{\mathrm{b}})$ represents the balance between the desolvation energy required to insert $Li^+$ into the channel and the electrochemical work supplied by the external circuit.

_Li-Li interactions._ Coulomb interactions between $Li^+$ ions are screened by both graphene and hBN. Using the method of image charges with multiple reflections, the effective interaction potential in reciprocal space is:

$$U_{\mathrm{eff}}(q) = U_0(q)\left[\frac{1 + (r_{\mathrm{gr}} + r_{\mathrm{hBN}})e^{-qd} + r_{\mathrm{gr}}\, r_{\mathrm{hBN}}\, e^{-2qd}}{1 - r_{\mathrm{gr}}\, r_{\mathrm{hBN}}\, e^{-2qd}}\right]. \qquad (2)$$

The bare Coulomb potential is $U_0(q) = e^2 (4\pi\varepsilon_0)^{-1}\, 2\pi q^{-1}$, with $q$ the in-plane wave vector of electrons in graphene. The reflection coefficients are $r_{gr}(q) = -q_{TF}/(q+q_{TF})$ and $r_{hBN} = -(\varepsilon_{hBN}-1)/(\varepsilon_{hBN}+1)$, with Thomas-Fermi screening vector $q_{TF} = ge^2 k_F\, (4\pi\varepsilon_0 \hbar v_F)^{-1}$ and Fermi wave vector in graphene $k_F = \sqrt{(\pi n)}$.The real-space interaction is obtained using the Hankel transform: $U_{\mathrm{eff}}(z) = \frac{1}{2\pi}\int_0^\infty J_0(qz)\, U_{\mathrm{eff}}(q)\, q\, dq$, where $J_0(qz)$ is the first-kind Bessel function. The Li-Li interaction energy per surface area is thus:

$$E_{\mathrm{Li-Li}} = n_{\mathrm{Li}} U_{\mathrm{eff}}(z) \qquad (3)$$

with $z = \sqrt{(\pi\, n_{\mathrm{Li}})}$, which depends on the electronic density in the system and thus on the applied gate voltages. Typically, we have $n_{Li} = 10^{13} - 10^{14}\ \mathrm{cm}^{-2}$, which gives an average Li – Li separation $r$ = 1.8 – 0.86 nm, much larger than the channel width $d \approx 0.35$ nm, justifying that we treat the $Li^+$ layer as a dilute 2D Coulomb gas confined between two screening planes.

_Li-host interaction._ Each $Li^+$ ion interacts with its own image charge on both graphene and hBN. This corresponding self-energy is obtained from the $z \to 0$ limit of the screened potential: $U_{\mathrm{self}}(q_{\mathrm{TF}}) = \frac{e^2}{4\pi\varepsilon_0}\int_0^\infty \left(\frac{U_{\mathrm{eff}}(q)}{U_0(q)} - 1\right) dq$. The Li–host interaction energy per unit area is therefore

$$E_{\mathrm{Li-host}} = n_{\mathrm{Li}} U_{\mathrm{self}}. \qquad (4)$$

_Insertion energy balance._ Insertion of $Li^+$ from the electrolyte into the low-dielectric channel incurs a desolvation penalty. In the Born approximation[9-11]: $\Delta G_{\mathrm{desolvation}} = -\Delta G_{\mathrm{solvation}} = +\frac{ze^2}{8\pi\varepsilon_0 R_{\mathrm{eff}}}\left[1 - \frac{1}{\varepsilon_r}\right]$, with $R_{\mathrm{eff}}$ the effective Born radius. Since $\varepsilon_r >> 1$, this reduces to:

$$\Delta G_{\mathrm{desolvation}} \approx a_{\mathrm{Born}} = \frac{e^2}{8\pi\varepsilon_0 R_{\mathrm{eff}}}. \qquad (5)$$

Using $R_{\mathrm{eff}} \approx 0.13$ nm, yields $a_{Born} \approx 5.5$ eV. The electrochemical work supplied by the circuit is $eV_{\mathrm{b}}$, where $V_b = (\Sigma V + \Delta V)/2$, so the intercalation threshold corresponds to a line in the $(\Sigma V, \Delta V)$–plane, given by $\Sigma V = 2V_{\mathrm{b}} - \Delta V$.

*Discrete intercalation states*. The experiments show that only four discrete $Li^{+}$ densities are allowed: $n_{Li} = l \times \Delta n_{Li}$, with $\Delta n_{Li} \approx 2 \times 10^{13}$ $cm^{-2}$ and $l = 0, 1, 2, 3$. For each density, the Gibbs energy is calculated using Eq. (1). Fig. S10 shows the resulting $G(l)$, which predict the stable density states as gate voltages are varied.

*Ionic Current.* The ionic current is described by a drift-diffusion model. For the radial geometry of the device: $2 \pi \ln(r_2/r_1)^{-1} e n_{Li} \mu_{Li} \Delta V \exp(-E_a/k_B T)$. The model predicts that the current will increase in discrete steps as the system transitions through the states $l = 1, 2, 3$ and will exhibit hysteresis, mirroring the behaviour of $n_{Li}$.

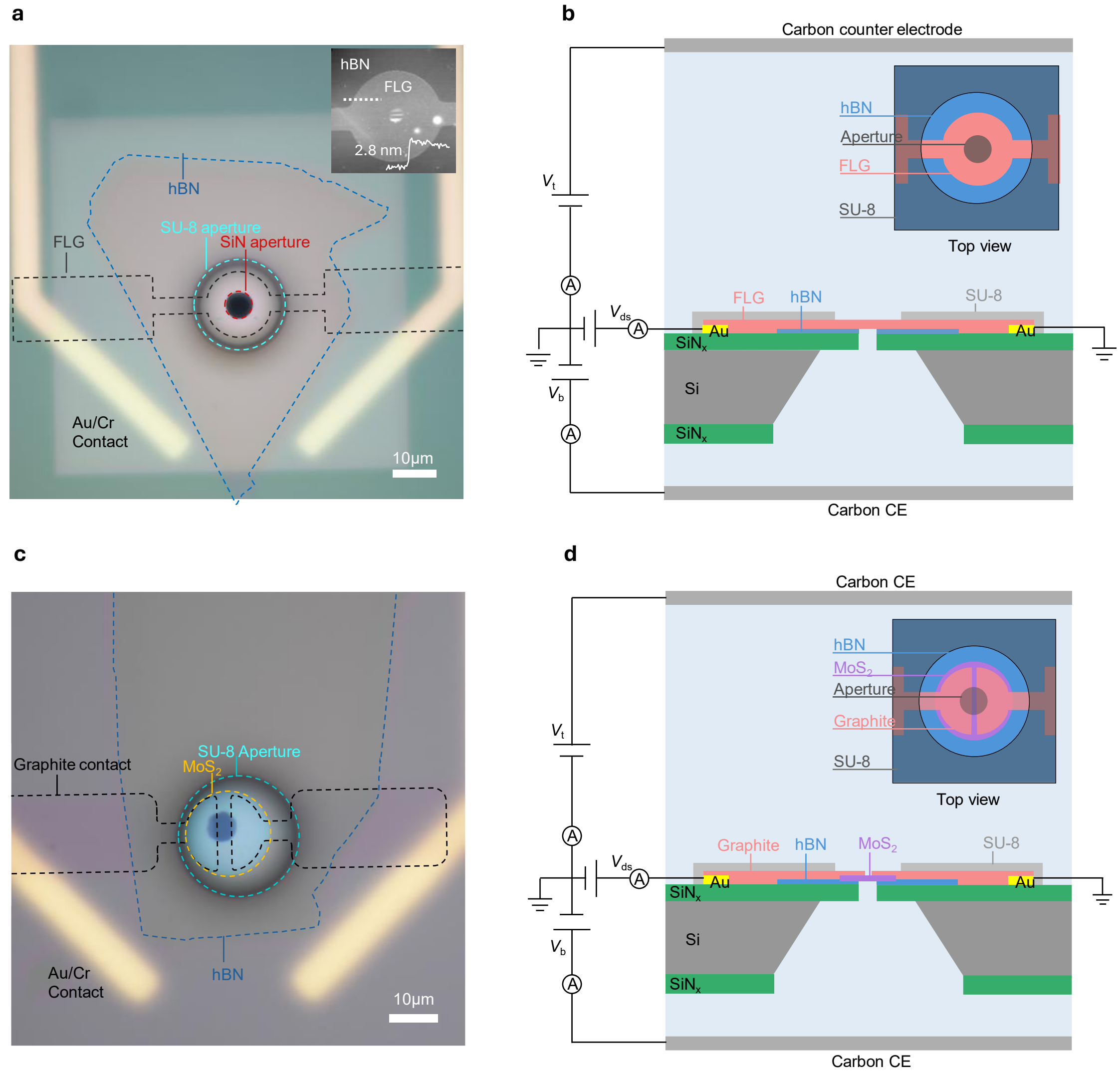

**Fig. S1. Device images and schematics**. **a**, Optical image of a typical FLG device. Dashed blue lines mark the hBN flake and dotted black lines mark the patterned FLG flake, which is connected to the Au/Cr contacts. The dashed cyan circle marks the aperture in the SU-8 washer. The entire field of view, except for the aperture in the SU-8 washer, is covered with SU-8. The dashed red line marks the aperture etched in SiN and hBN. Inset, AFM image of the FLG/hBN device, with the corresponding height profile along the dashed line shown beneath. **b**, Cross-section schematic of the FLG device with electrical connections. Inset, top view schematic. **c**, Optical image of a $MoS_2$ device. hBN crystal and SU-8 clamp are outlined as in **a**. The yellow dashed circle marks patterned $MoS_2$ crystal, the dashed black line outlines graphite contacts that electrically connect the $MoS_2$ flake to the Au/Cr contacts. The gap between graphite electrodes ensures the electric resistance is limited by the $MoS_2$ flake. **d**, Cross-section schematic of the device shown in panel **c**. Inset, top view schematic.

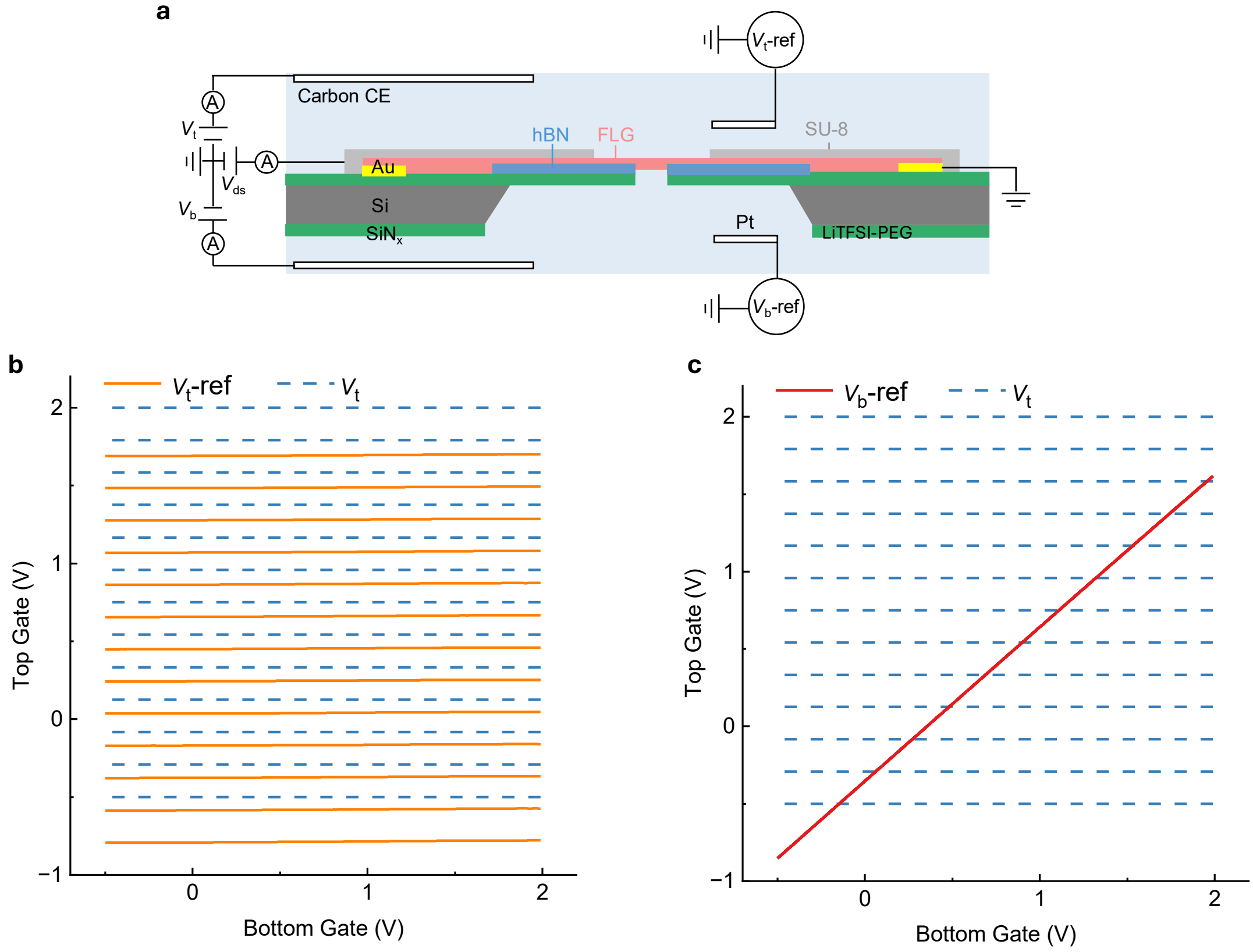

**Fig. S2. Independence of top and bottom gates**. **a**, Cross-section schematic of a FLG device with electrical connections and Pt pseudo-reference electrodes placed in the top and bottom reservoirs, which are measured simultaneously with the gate voltage. **b**, Voltage measured with a Pt pseudo-reference electrode located in the top electrolyte reservoir ($V_t$-ref, orange) as a function of $V_b$ for fixed $V_t$ (dashed blue lines). $V_t$-ref is unaffected by the applied $V_b$. $V_t$-ref and $V_t$ display a constant offset of 0.4 V. **c**, Voltage measured with a Pt pseudo-reference electrode located in the bottom electrolyte reservoir ($V_b$-ref, red) under the same conditions as in panel **b**. The measurements reveal that $V_b$-ref varies linearly with $V_b$ and is unaffected by changes in $V_t$. The two measurements demonstrate that the top and bottom potentials are independent.

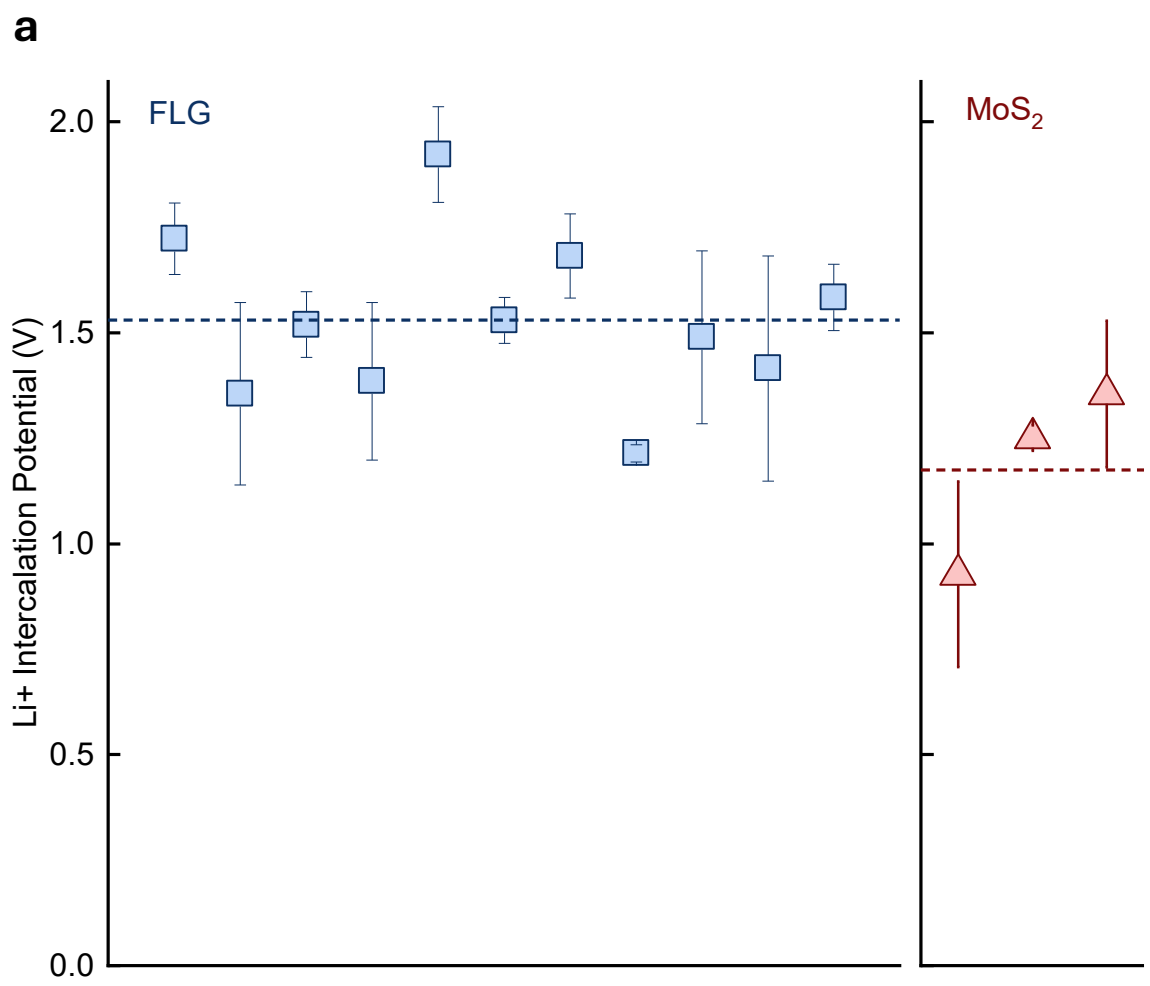

**Fig. S3. Threshold potentials for intercalation in FLG and $MoS_2$ devices**. Scatter plot showing the distribution of threshold potentials across 11 FLG (blue squares) and 3 $MoS_2$ (red triangles) devices. Each point represents the mean across multiple scans; error bars are the standard deviation. Dashed blue and red lines, mean threshold potential of FLG and $MoS_2$ devices, respectively. The mean turn-on potential for FLG devices is 1.5 ± 0.1 V; for $MoS_2$ devices it is 1.2 ± 0.1 V.

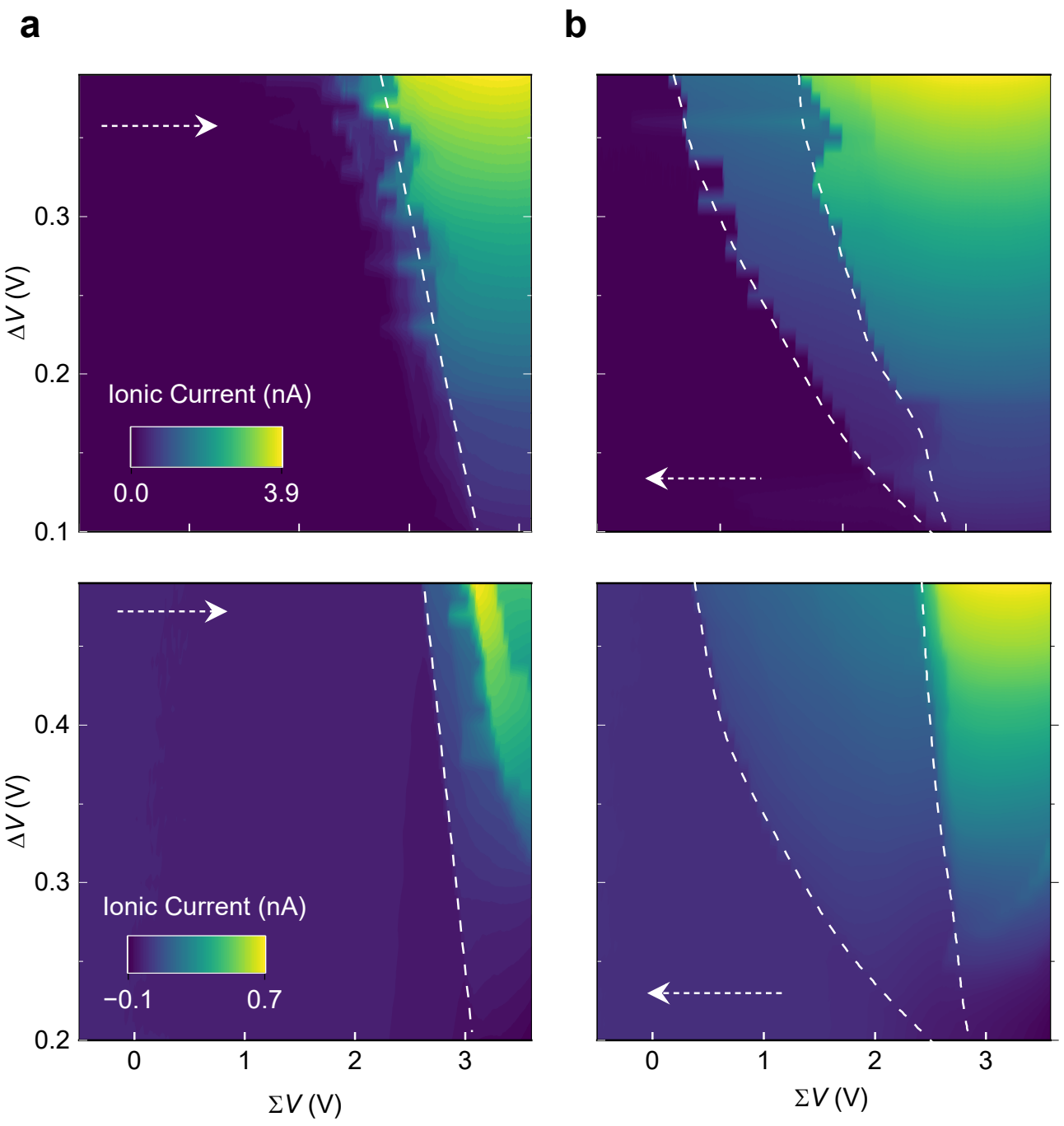

**Fig. S4. Ionic transport maps from additional devices.** Forward scans (panel **a**) and backward scans (panel **b**) from two additional FLG devices (top and bottom panels). Both devices show the same qualitative set of features displayed in Fig. 2, namely $V_b$-dependent intercalation and discharge features with non-linear dependence on $\Sigma V$ and $\Delta V$. Arrows denote scan direction. Dashed lines, guides to the eye.

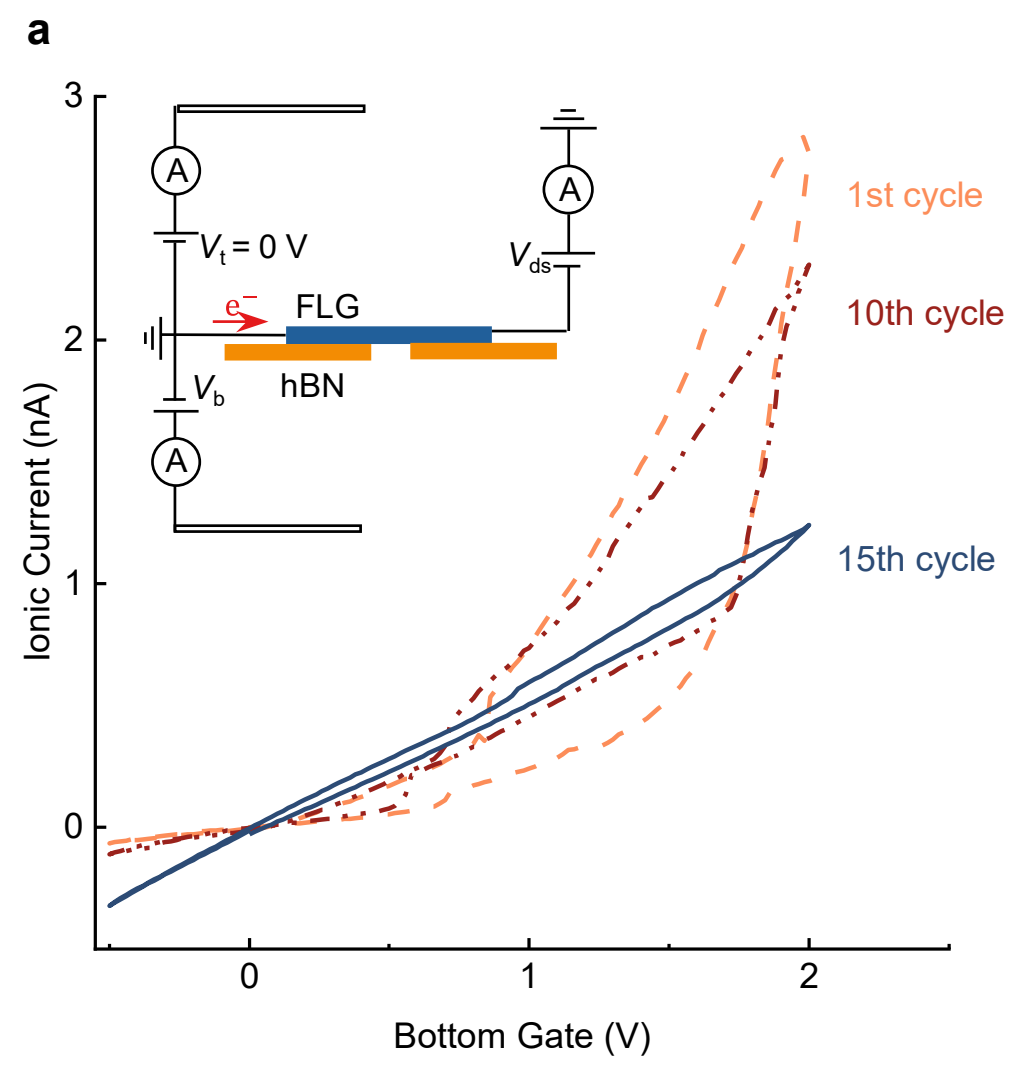

**Fig. S5. Devices measured with a single gate voltage. a**, Ionic current as a function of a single gate voltage (the opposite gate held at zero volts). Inset, schematic of single-gate configuration. The configuration is identical to that of Fig.1 a, only with fixed $V_t$ = 0 V. Successive traces were obtained by repeatedly cycling the bottom gate voltage. The initial sweeps show hysteresis, which disappears within a few cycles as the device degrades, resulting in linear traces indicative of irreversible leakage.

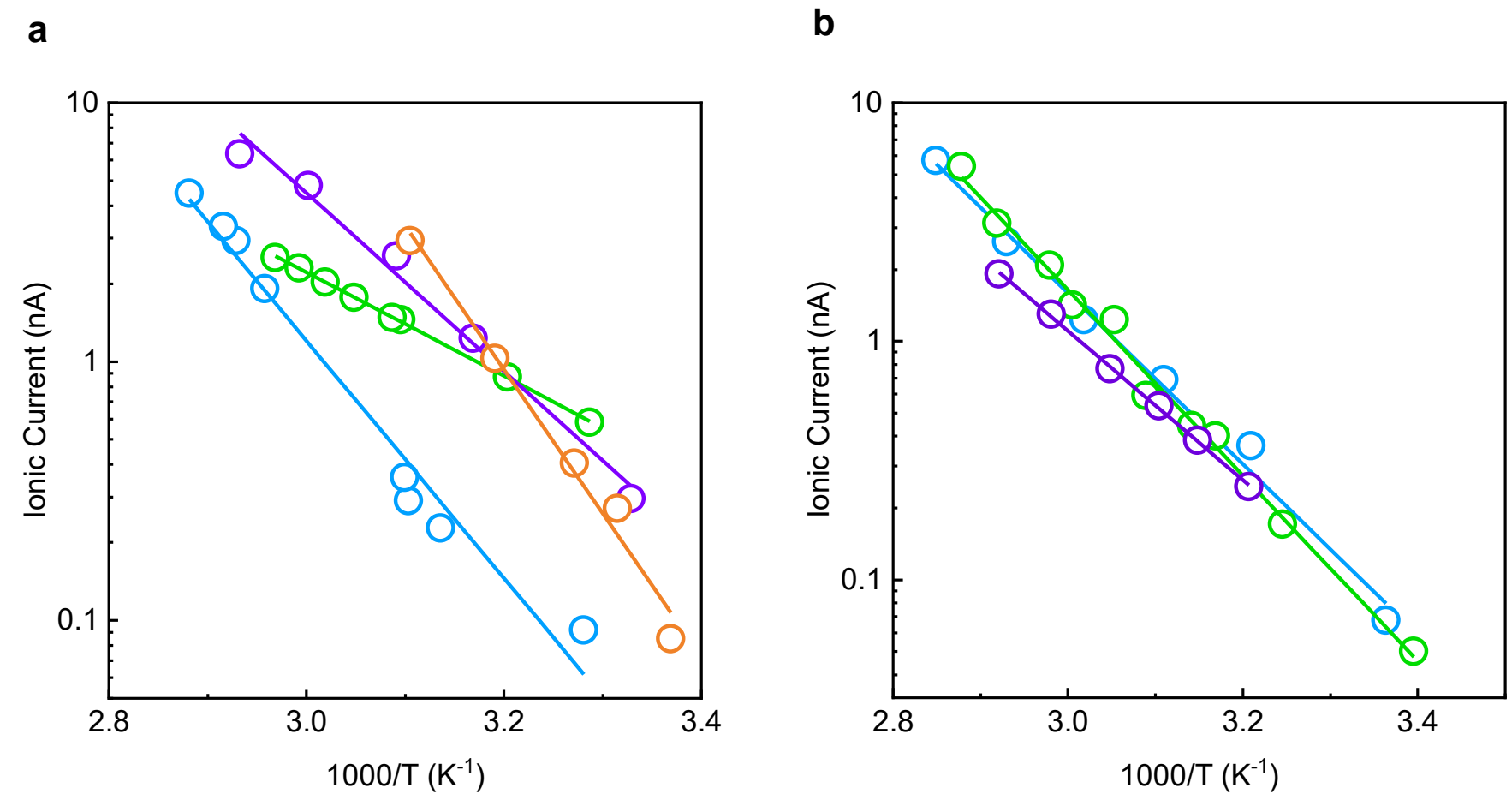

**Fig. S6. Temperature dependence of $Li^+$ transport in double-gated van der Waals devices.** Arrhenius plots of ionic current from (**a**) FLG and (**b**) $MoS_2$ devices. Each data point is the average current at the highest plateau in the current traces. Each colour marks data from a different device. Solid lines, best fits to data, from which we extract $E_a = 0.6 \pm 0.2$ eV for FLG and $E_a = 0.7 \pm 0.1$ eV for $MoS_2$ from the different devices.

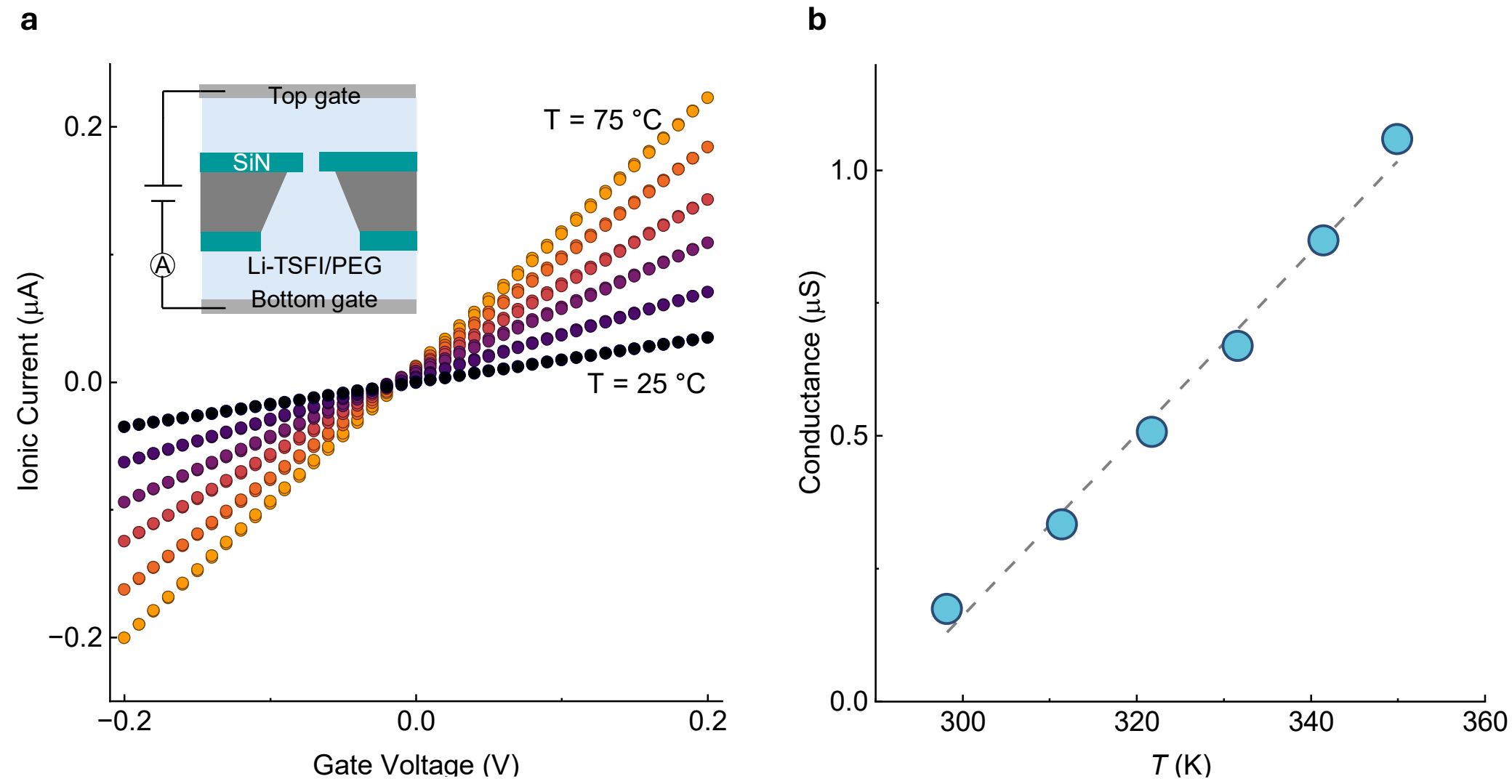

**Fig. S7. Conductance and temperature dependence of bulk electrolyte. a**, Current–voltage response of 'open-hole' devices consisting of a 5-μm-diameter-hole etched in a silicon nitride substrate. At all temperatures, the current-voltage response is linear, and the current magnitude is several orders of magnitude higher than that measured for FLG and $MoS_2$ devices. Inset, schematic of the open-hole device used to measure conductance. **b**, Electrolyte conductance plotted as a function of temperature. Dashed black line, best linear fit to data.

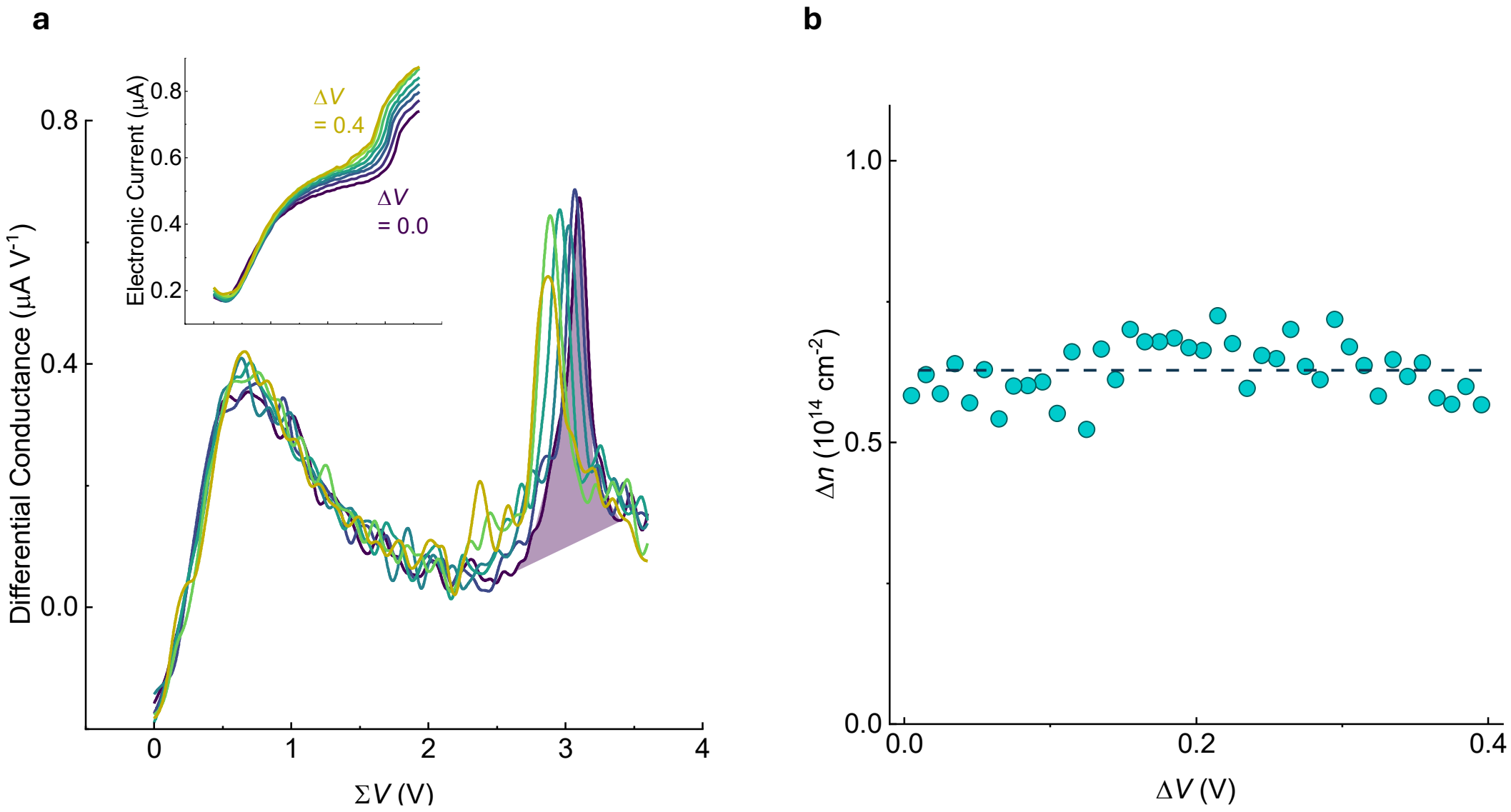

**Fig. S8. Estimation of *Δn* from electronic differential conductance. a**, Derivative of the in-plane electronic conductance (non-background-subtracted) as a function of $\Sigma V$ for $\Delta V$ = 0–0.4 V. Purple shaded region is the baseline-subtracted area of the peak corresponding to the intercalation of $Li^+$ ions. Inset, raw data of in-plane electronic conductance traces under the same conditions as the main panel. At low potentials, the signal shows typical gating behaviour for FLG, with the Dirac point at $\Sigma V \approx 0.2$ V. A stepwise change is observed around $\Sigma V = 3$ V. **b**, Electronic charge density extracted by integrating the peaks in the differential conductance for different $\Delta V$, as shown in panel **a**. The dashed line marks the average charge density.

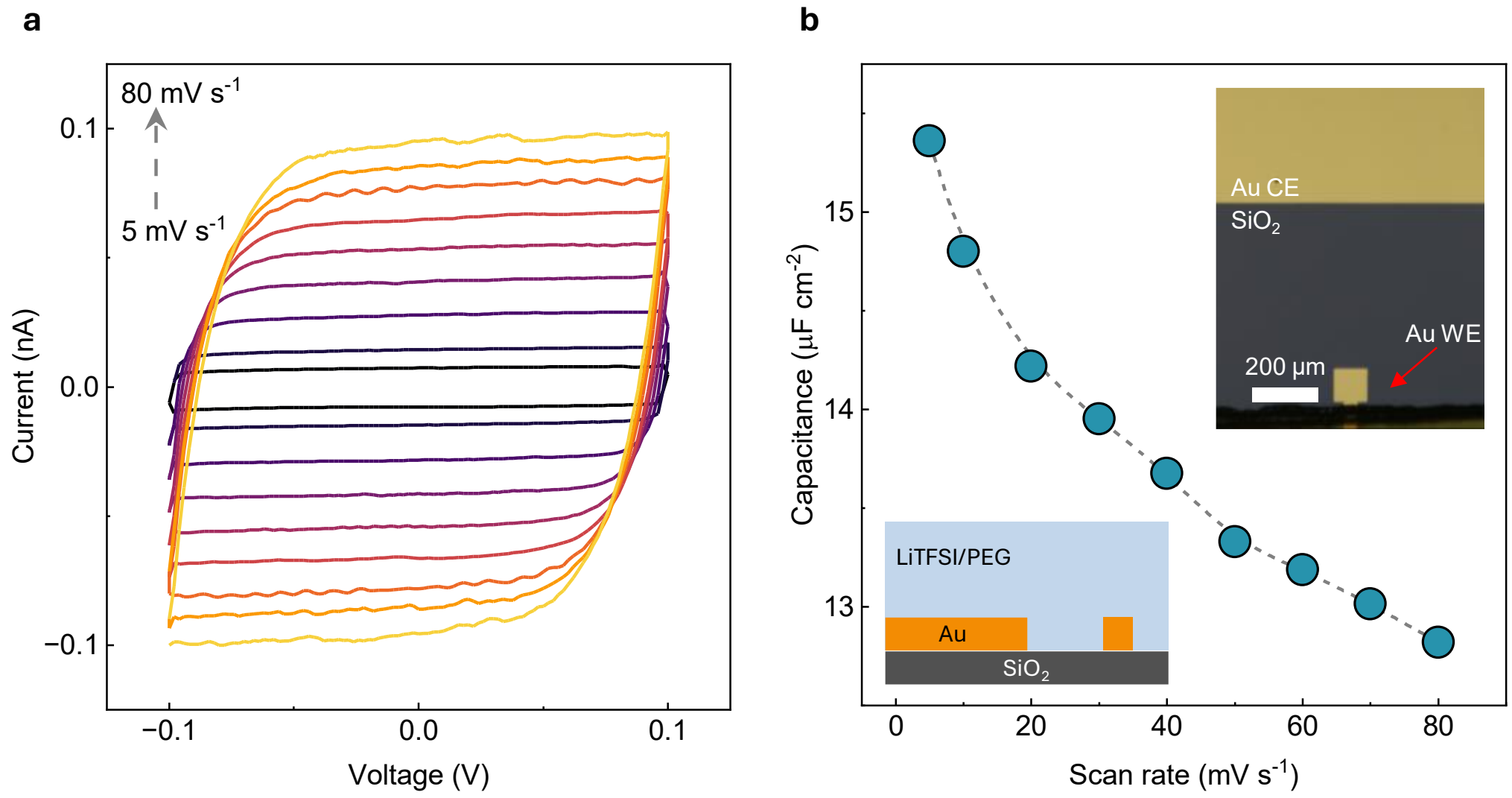

**Fig. S9. Electrolyte capacitance**. **a**, Cyclic voltammetry curves from reference devices consisting of two Au electrodes in contact with the LiTFSI electrolyte used throughout this work. The voltage is scanned from −0.1 to 0.1 V at rates from 5–80 mV $s^{-1}$. **b**, Capacitance per unit area extracted from **a**. Top inset, optical image of the device. Bottom inset, cross section schematic of the device.

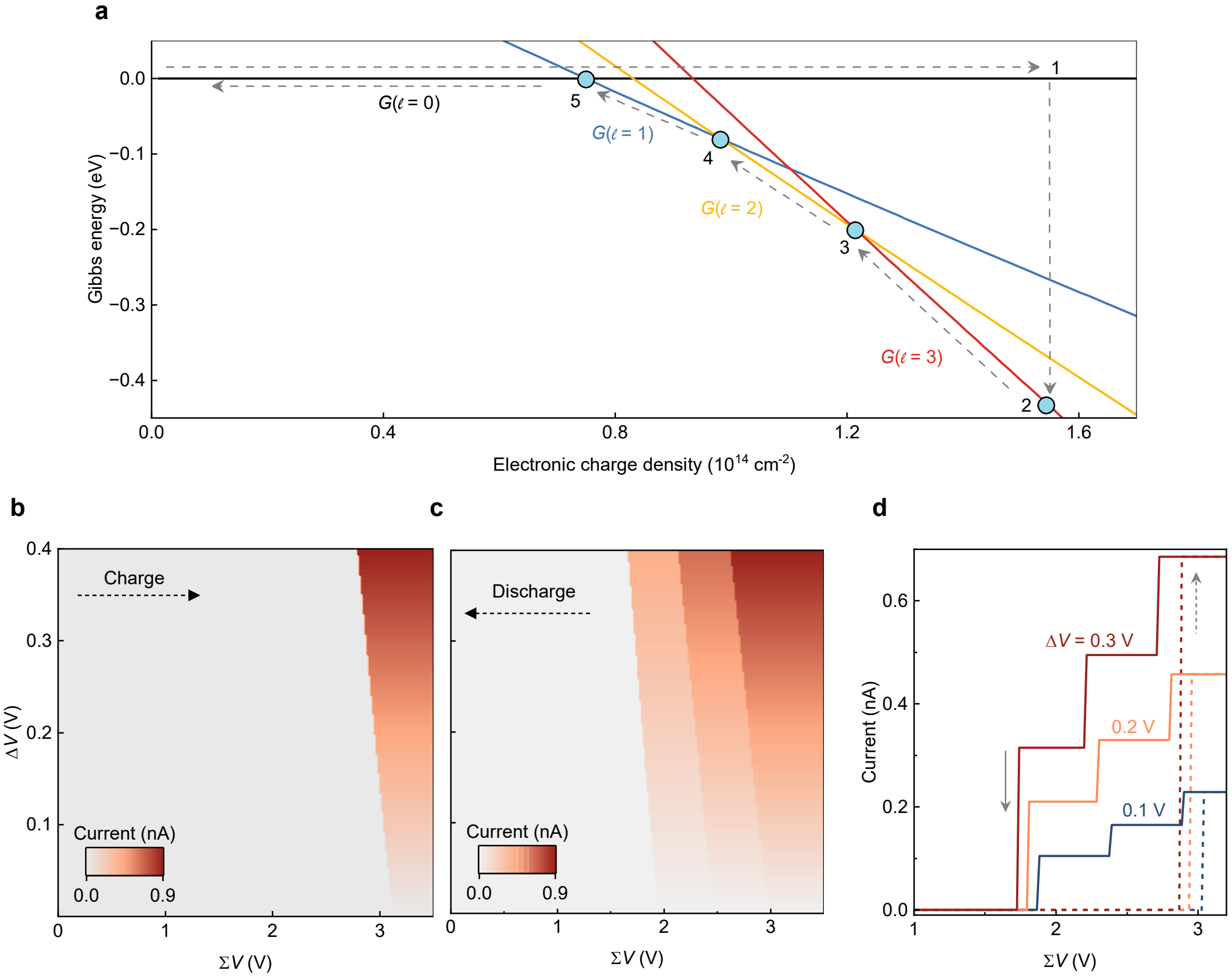

**Fig. S10. Analytical model. a**, Gibbs free energy as a function of electronic density in the channel for the four allowed ionic densities, $n_{Li} = \ell \times \Delta n_{Li}$, with $\Delta n_{Li} \approx 2 \times 10^{13}$ cm$^{-2}$ and $\ell = 0, 1, 2, 3$ Each discrete ionic density defines a distinct free-energy branch, G($\ell$), shown in different colours. Dashed arrows indicate intercalation and deintercalation trajectories along each branch, while blue dots mark transition points between adjacent G($\ell$) branches. **b-c**, Ionic current maps, calculated from the analytically derived $n_{Li}$ using a drift-diffusion transport model (Section S7). Maps exhibit qualitative features similar to that of experiment, with hysteretic charge-discharge behaviour and discharge plateaus. **d**, Traces at constant $\Delta V$ obtained from panels **b** and **c**. Arrows mark sweep direction. Dashed lines, charging sweep. Solid lines, discharging sweep.

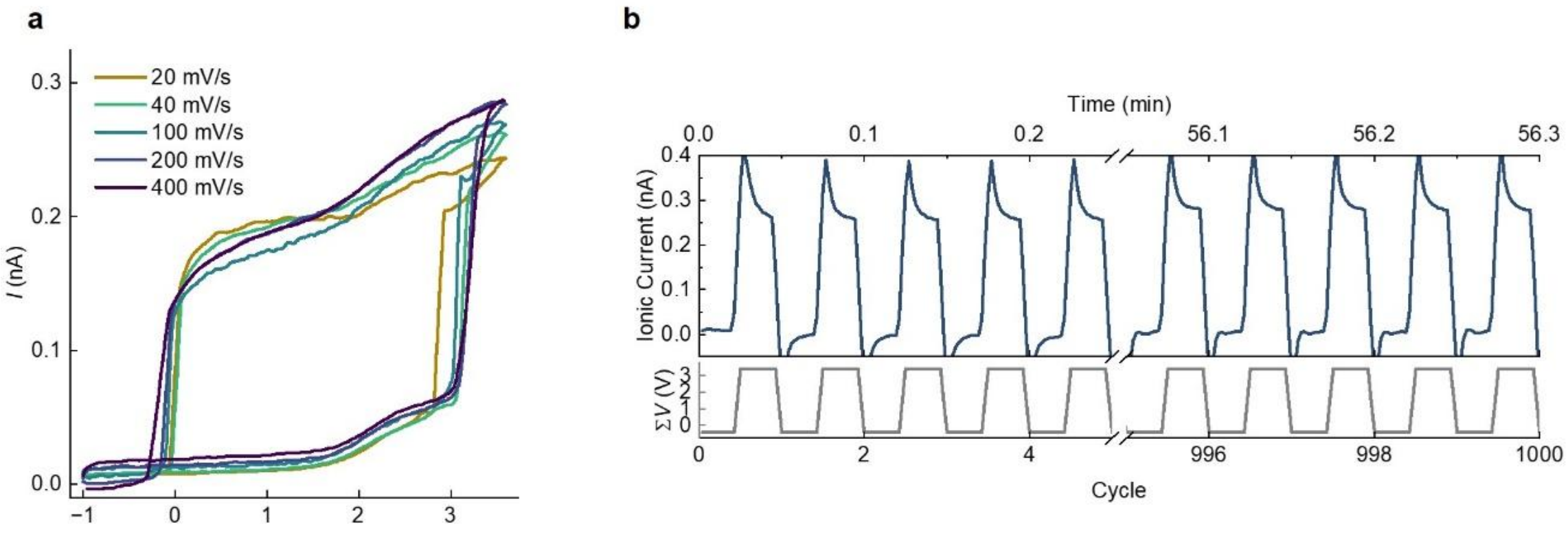

**Fig. S11. | Scan-rate and cycling-rate dependence. a,** Ionic current as a function of Σ*V* for scan rates from 20 to 400 mV $s^{-1}$. Hysteresis is preserved across scan rates, indicating that the steady-state response is not dominated by capacitive charging. **b,** Square-wave cycling of the ionic transport signal for 1000 cycles at 300 mHz. The response remains stable and reproducible over extended cycling. Small transient features immediately after switching are attributed to capacitive charging, whereas the persistent plateau current reflects steady-state $Li^+$ transport.

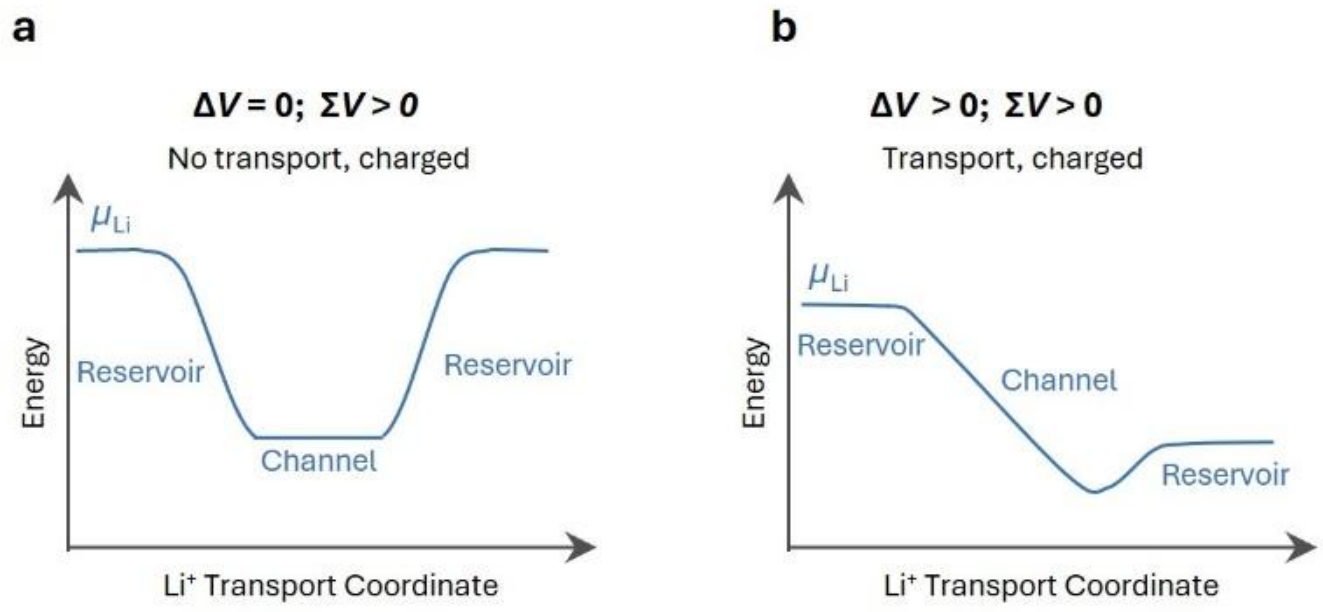

**Fig. S12 | Energy profile of $Li^+$ in double-gated devices.** Schematic potential landscape along the $Li^+$ channel and reservoirs. **a**, The common-mode gate voltage $\Sigma V = V_t + V_b$ shifts the electrochemical potentials of both reservoirs together, creating favourable conditions for ion insertion. **b**, The differential gate tilts the energy profile of the channel, inducing net ion flow.